\documentclass[prx,aps,twocolumn,superscriptaddress,nofootinbib]{revtex4}
\usepackage{graphicx}
\usepackage{dcolumn}
\usepackage{bm}
\usepackage{amsmath}
\usepackage{epsfig}
\usepackage{color}
\usepackage{soul}

\newcommand{\abs}[1]{\left|#1\right|}

\newcommand{\R}[1]{\textrm{Re}\{#1\}}
\newcommand{\mean}[1]{\ensuremath{\left\langle #1 \right\rangle}}
\newcommand{\bra}[1]{\ensuremath{\left\langle #1 \right|}}
\newcommand{\ket}[1]{\ensuremath{\left| #1 \right\rangle}}

\def\f12{\frac{1}{2}}

\long\def\symbolfootnote[#1]#2{\begingroup%
 \def\thefootnote{\fnsymbol{footnote}}\footnotetext[#1]{#2}\endgroup}

\def\Ab{{\rm \bf A}}   \def\Eb{{\rm \bf E}}
  \def\Jb{{\rm \bf J}} \def\Kb{{\rm \bf K}} \def\Pb{{\rm \bf P}}
\def\db{{\rm \bf d}} \def\gb{{\rm \bf g}} \def\jb{{\rm \bf j}} \def\kb{{\rm \bf k}} \def\qb{{\rm \bf q}}
\def\rb{{\rm \bf r}}
\def\Jdc{J_{\rm DC}}       \def\Jbac{\Jb_{\rm AC}}
 \def\jbdc{\jb_{\rm DC}}  \def\jbac{\jb_{\rm AC}}
\def\kpq{{\kb+\qb}} 

\def\ii{{\rm  i}}   \def\ee{{\rm  e}} \def\dd{{\rm  d}}
 
\def\Hi{{\hat{H}_{\rm I}}}
\def\pb{{\hat{\rm \bf p}}}
\def\den{{\hat{\rho}}}
 \def\me{{m_{\rm e}}} \def\Ef{{E_{\rm  F}}}
\def\kf{{k_{\rm  F}}}
 
\def\nef{{n_{\rm eff}}}
\def\gM{{g_{\rm M}}}  \def\gA{{g_{\rm A}}}
\def\gt{{g_{\rm tot}}}  \def\gf{{g_{\rm EM}}}  \def\gm{{g_{\rm mat}}}  \def\ge{{g_{\rm ele}}}  \def\gl{{g_{\rm lat}}}
            \def\gbe{{\gb_{\rm ele}}}             
\def\gbm{{\gb_{\rm mat}}}           

\def\nomde{{\omega_{j'\kpq}-\omega_{j\kb}-\omega}}
\def\nfdif{{f_{j'\kpq}-f_{j\kb}}}
\def\jk{{j\kb}} \def\jpkq{{j'\kpq}}

\begin{document}
\title{Emergence of material momentum in optical media}
\author{Deng~Pan}
\email[Corresponding author: ]{Deng.Pan@icfo.eu}
\affiliation{ICFO-Institut de Ciencies Fotoniques, The Barcelona Institute of Science and Technology, 08860 Castelldefels (Barcelona), Spain}
\author{Hongxing~Xu}
\affiliation{School of Physics and Technology, Wuhan University, Wuhan 430072, China}
\author{F.~Javier~Garc\'{\i}a~de~Abajo}
\affiliation{ICFO-Institut de Ciencies Fotoniques, The Barcelona Institute of Science and Technology, 08860 Castelldefels (Barcelona), Spain}
\affiliation{ICREA-Instituci\'o Catalana de Recerca i Estudis Avan\c{c}ats, Passeig Llu\'{\i}s Companys 23, 08010 Barcelona, Spain}


\begin{abstract}
Understanding the momentum of light when propagating through optical media is not only fundamental for studies as varied as classical electrodynamics and polaritonics in condensed matter physics, but also for important applications such as optical-force manipulations and photovoltaics. From a microscopic perspective, an optical medium is in fact a complex system that can split the light momentum into the electromagnetic field, as well as the material electrons and the ionic lattice. Here, we disentangle the partition of momentum associated with light propagation in optical media, and develop a quantum theory to explicitly calculate its distribution. The material momentum here revealed, which is distributed among electrons and ionic lattice, leads to the prediction of unexpected phenomena. In particular, the electron momentum manifests through an intrinsic DC current, and strikingly, we find that under certain conditions this current can be along the photonic wave vector, implying an optical pulling effect on the electrons. Likewise, an optical pulling effect on the lattice can also be observed, such as in graphene during plasmon propagation. We also predict the emergence of boundary electric dipoles associated with light transmission through finite media, offering a microscopic explanation of optical pressure on material boundaries.
\end{abstract}
\date{\today}

\maketitle

\section{Introduction}

Understanding the momentum of light when propagating through optical media is important for both fundamental and applied perspectives. In history, shortly after the formulation of Maxwell's equations, two possible solutions were proposed by Minkowski and Abraham for the momentum of light propagating in a medium of refractive index $n$, which in modern language correspond to respectively $n\hbar q_0$ and $\hbar q_0/n$ per photon, where $q_0$ is the free-space wave vector. These solutions gave birth to a controversy that has been the subject of heated debate \cite{Milonni2010} from both theoretical \cite{Gordon1973,Nelson1991,Leonhardt2006,Pfeifer2007,Barnett2010,Milonni2010,Mansuripur2010,Kemp2011,Griffiths2012,Silveirinha2017,Partanen2017} and experimental \cite{Jones1954,Ashkin1973,Walker1975,Jones1978,Campbell2005,She2008,Astrath2014} perspectives. Various theoretical studies have reached a consensus \cite{Barnett2010,Mansuripur2010,Kemp2011,Pfeifer2007,Kemp2011,Griffiths2012,Partanen2017}, suggesting that the Abraham form stands for the electromagnetic (EM) field momentum, while the Minkowski form includes the material momentum.

Nowadays, the importance of having in-depth knowledge on light momentum in material media has extended beyond classical electrodynamics and its related application in optical manipulation \cite{Gao2017} to various frontiers in condensed matter and quantum physics, such as new photovoltaic phenomena \cite{Vengurlekar2005,Strait2019,Karch2010}, optically-induced magnetization \cite{Bliokh2017,Bliokh2017_2,Rudner2019}, photon-coupled polariton states \cite{Hwang2007,Yoxall2015,Lundeberg2017,Hu2017,Basov2016}, or even quantum-driven systems \cite{Thouless1983,Wang2013,Privitera2018,Eckardt2017,Cheng2020}. From a microscopic perspective, an optical medium is in fact a complex system, where the material momentum discussed in the AM controversy can be further decomposed into material electrons and lattice ions contributions, thus requiring a detailed account of the electronic band structure. As a consequence of the momentum in the material electrons, a photovoltaic effect should be generated even in a dissipationless medium, as manifesting in the photon-drag effect observed under oblique light incidence on a thin metallic film \cite{Vengurlekar2005,Strait2019} or graphene \cite{Karch2010}, which differs from the photogalvanic effect \cite{Gibson1970,Danishevskii1970,Luryi1987,Wieck1990} originating in the highly absorptive interband electron-hole pair excitations. Very unexpectedly, a recent photon-drag experiment \cite{Strait2019} reports that even the direction of the electric current excited in a metallic film is at odds with respect to the widely-accepted theoretical prediction, therefore revealing the complexity of the momentum transfer process in such a seemingly simple optical medium. Benefiting those broad topics pertinent to light momentum in media, it is thus fundamentally important to reconsider this old problem from first principles using microscopic quantum theory.

In this article, we formulate a general quantum theory to disentangle the momentum distribution associated with light propagation in various types of media. We explicitly calculate the momentum distribution in the EM field, electrons, and ionic lattice, which leads to a series of inspiring results. More precisely, (1) for transverse EM waves, our explicit result of the ratio between the field and material (electrons plus lattice) obtained based on quantum theory is consistent with previous classical solutions to the AM problem. (2) We conclude that a nonzero electron momentum generally exists in media with partially filled electron bands, which manifests as a DC current. Interestingly, this DC current is also observed due to interband transitions for a photon energy lower than the band gap energy. (3) Surprisingly, in some scenarios electrons are pulled back by an incident light wave, thus shedding light into recent counter-intuitive experimental observations \cite{Strait2019}. A similar pulling effect on the lattice should be equally observed in graphene during plasmon propagation. (4) We also predict the emergence of boundary electric dipole moments during light transmission through finite insulating media, which provide a new microscopic interpretation of optical pressure.

\begin{figure*}
\centering
\includegraphics[width=1\textwidth]{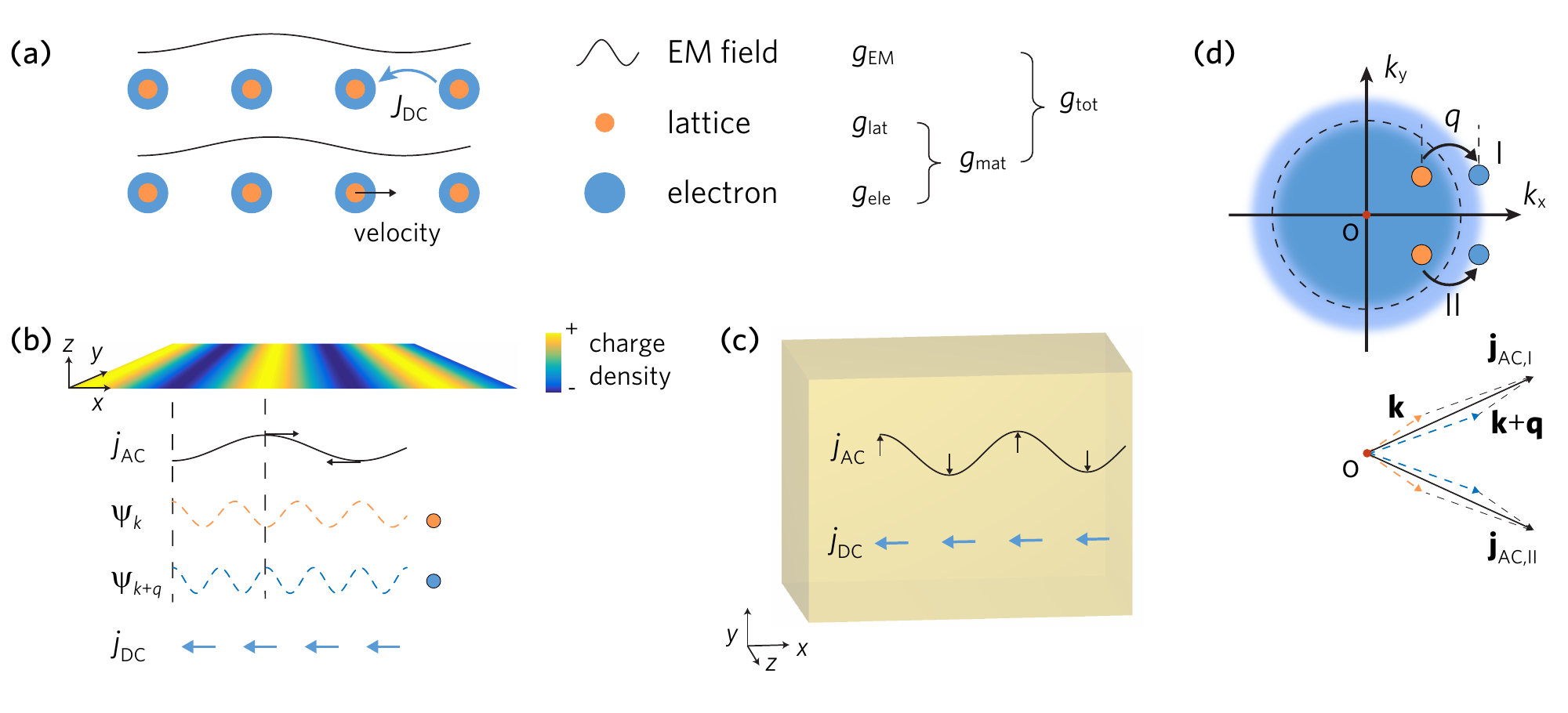}
\caption{ Illustration of light propagation in infinite optical media. (a) The total momentum $\gt$ associated with light propagation in a medium is distributed in the field ($\gf$), the lattice ($\gl$) and the electrons ($\ge$). The electron and lattice momenta, manifesting in a DC current (blue arrow) and a lattice velocity (black arrow), respectively, form the material momentum $\gm$. (b) Plasmons on a 2DEG and (c) photons in a bulk dielectric medium support longitudinal and transverse AC current-density waves (black curves and arrows), respectively. In both scenarios, $\ge$ can result in DC currents (blue arrows). (d) Inside a medium, the EM field is combined with electronic quantum transitions between states denoted by color-filled circles in the momentum space representation of the upper panel. The transitions I and II produce AC current waves $\jb_{\rm AC,I}$ and $\jb_{\rm AC,II}$ (see Eq.\ (\ref{jac})) whose directions are shown by black arrows in the lower panel. The in-phase and out-of-phase superpositions of $\jb_{\rm AC,I}$ and $\jb_{\rm AC,II}$ transition currents result in the total longitudinal and transverse AC current-density perturbances observed in plasmons and transverse EM waves shown in (a) and (b), respectively. In plasmons, the charge-density waves (color scale in (b)) are formed by the quantum superposition of pairs of electronic quantum states $\psi_k$ and $\psi_{k+q}$ of different momenta (dashed curves; for illustration purposes both momenta are assumed to be along $\hat{x}$). The resulting real-space charge-density $-e\abs{\Psi(x,t)}^2$ is presented in the color scale. During the transitions shown in (d), both the electrons and lattice gain net momenta of $\ge$ and $\gl$ [see Eqs.\ (\ref{gT}) and (\ref{gL})], as illustrated in (a).
}
\label{Fig1}
\end{figure*}

\section{Results and discussion}

\subsection{Momentum decomposition in optical media}

We consider here an EM plane wave propagation in a lossless and infinitely extended optical medium that is composed of electrons and a crystal lattice, as illustrated in Fig.\ \ref{Fig1}(a). Two typical scenarios for light propagation in optical media are also shown in Fig.\ \ref{Fig1}(b) and (c): a plasmon polariton wave sustained by a two-dimensional electron gas (2DEG) [Fig.\ \ref{Fig1}(b)], and a transverse EM plane wave propagating in a 3D bulk transparent dielectric medium [Fig.\ \ref{Fig1}(c)]. Unless otherwise stated, we chose a reference frame in which the medium is in thermal equilibrium and the momentum of both electrons and lattice are zero before applying any light field. After introducing a monochromatic light field propagating in such a medium and eventually reaching a steady state, as illustrated in Fig.\ \ref{Fig1}(a), the total momentum in the system $\gt$ should include components distributed in the field ($\gf$), the electrons ($\ge$) and the lattice ($\gl$), where the sum $\gt=\ge+\gl$ is the material momentum on which we focus in this study. We note that both $\ge$ and $\gl$ are indeed physical observables, which manifest in a DC current and the classical velocity of the lattice, respectively [see Fig.\ \ref{Fig1}(a)]. To describe the emergence of these types of momenta associated with matter, a microscopic quantum theory becomes necessary, which should go beyond classical theory and lead to explicit results of $\ge$ and $\gl$. Disentangling these types of momenta allows us to reveals new optomechanical and photoelectric phenomena. In the following sections, we first revisit a microscopic explanation of the optical response of optical media, and then show our theoretical results for both $\ge$ and $\gt$ within the same framework.

\begin{figure*}[t]
\begin{centering}
\includegraphics[width=0.6\textwidth]{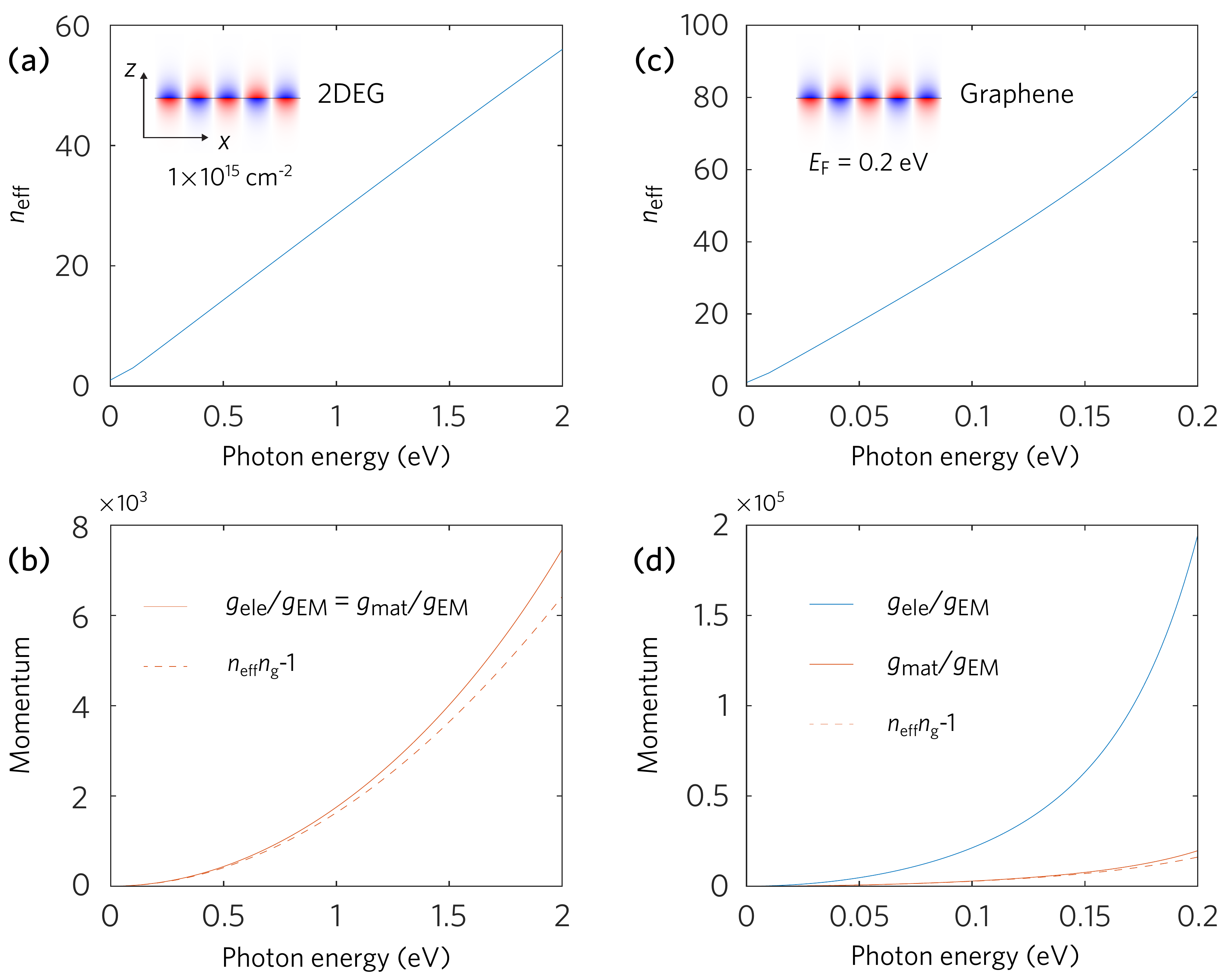}
\par\end{centering}
\caption{ Momentum distribution in 2D plasmons. (a) Effective refractive index and (b) momentum distribution of plasmons in a freestanding 2DEG [see carrier density in (a)]. In (b), the momenta distributed in electrons and ionic lattice, normalized to the field momentum $\gf$, are calculated based on our quantum model (solid curves, Eq.\ (\ref{gL})). For a 2DEG, the material momentum is entirely distributed in the electrons (i.e., $\ge=\gm$ and $\gl=0$). We can also estimate $\gm/\gf$ from the AM formalism (dashed curves). (c),(d) Same as (a) and (b) for monolayer graphene with Fermi energy $\Ef=0.2$\,eV. For graphene plasmons, the electron momentum $\ge$ is much larger than the material momentum $\gm$, so the lattice momentum $\gl$ should be antiparallel with the photonic wave vector.
}
\label{Fig2}
\end{figure*}

\begin{figure*}[t]
\begin{centering}
\includegraphics[width=1\textwidth]{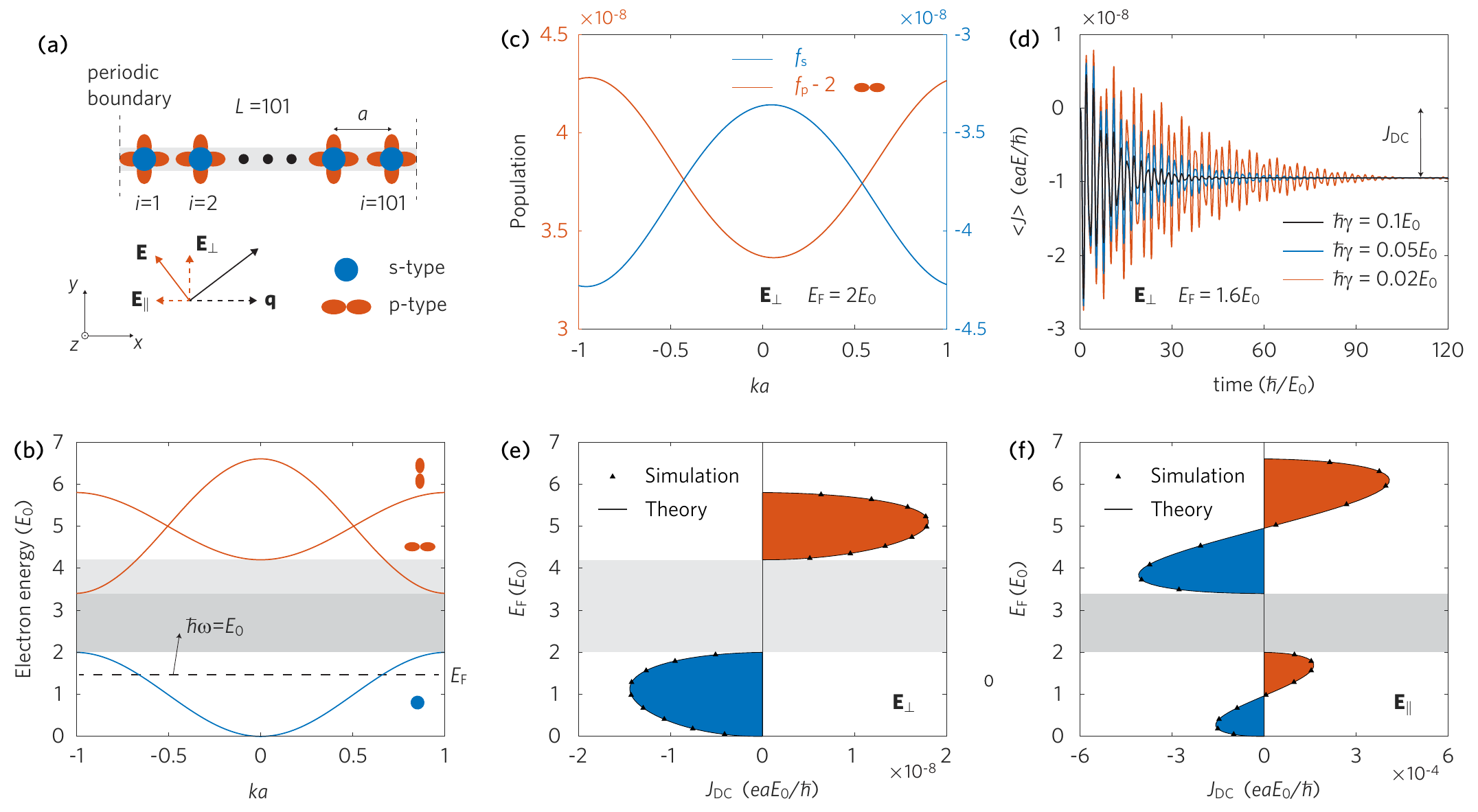}
\par\end{centering}
\caption{ Atomic chain under light excitation. (a) We consider an infinite atomic chain illuminated by an EM plane wave, where the optical responses to electric field components $\Eb_\perp$ and $\Eb_\parallel$ can be regarded as simplifications of Fig.\ \ref{Fig1} (b) and (c), respectively. The chain is modeled by $L=101$ atoms with periodic boundary conditions and neirest-neighbors binding couplings between the two lowest atomic orbitals, which are considered to be s- (blue) and p-type (red). The photonic wave vector along the chain is assumed to be $q=2\pi/La$. (b) Electronic band structure of the chain. The uncoupled s- and p-orbital energies are assumed to be $E_0$ and $5E_0$, expressed in terms of a characteristic energy $E_0$ of the system, with hopping energies of $0.5E_0$, $0.8E_0$ and $-0.4E_0$ for s-s, p$_\perp$-p$_\perp$ and p$_\parallel$-p$_\parallel$ couplings, respectively (written in units of $E_0$), which correspondingly result in s-, p$_\perp$- and p$_\parallel$-bands (see symbols on the right). (c) Steady populations in the s- and p$_\perp$-bands of an insulating chain ($\Ef=2E_0$) excited by $\Eb_\perp$. (d) Time evolution of the space-averaged current $\mean{J}$ in a chain with $\Ef=1.6 E_0$ excited by $\Eb_\perp$ for different damping frequencies $\gamma$. In (c) and (d), despite the assumption that the photon energy $\hbar\omega=E_0$ is smaller than the band gap energy, we still observe a population in the conduction band even in (c) an ideal insulator, and (d) a steady DC current $J_{\rm DC}$ in a chain with unfilled band. (e),(f) Dependence of $J_{\rm DC}$ on the Fermi energy $\Ef$ for electric field components $\Eb_\perp$ (e) and $\Eb_\parallel$ (f), calculated from Eqs.\ (\ref{gT}) and (\ref{gL}). In (c) to (f), a weak EM field is assumed (see Appendix). The polarization $\Eb_\perp$ in (c) to (e) can only induce interband transitions between s- and p$_\perp$-bands, while for $\Eb_\parallel$ intraband transitions are dominant, giving rise to plasmon waves. In either scenario, $J_{\rm DC}$ is oriented along $\qb_\parallel$ for certain range of $\Ef$ [red-filled areas in (e) and (f)], indicating an electron pulling effect.
}
\label{Fig3}
\end{figure*}

\subsection{Quantum description of optical media}

Assuming only linear optical response, the time-harmonic electric field $\Eb(\rb, t)=\R{\Eb(\omega)\ee^{\ii(\qb\cdot\rb-\omega t)}}$ of a monochromatic EM wave of frequency $\omega$ and wave vector $\qb$ induces a distribution of polarization vector $\Pb(\rb,t)=\R{\chi(\qb,\omega)\Eb(\omega)\ee^{\ii(\qb\cdot\rb-\omega t)}}$, dictated by the material's electric susceptibility $\chi(\qb,\omega)$, which consequently generates an alternating-current (AC) polarization density wave $\Jbac(\rb, t)=\partial_t \Pb(\rb, t)$. This wave is longitudinal in 2D plasmons and transverse in EM waves in a dielectric [black arrows and curves in Fig.\ \ref{Fig1}(a) and (c)]. Additionally, the longitudinal current $\Jbac(\rb, t)$ in 2D plasmons gives rise to a charge-density wave via charge continuity [color scale, Fig.\ \ref{Fig1}(b)]. The current-density wave $\Jbac(\rb, t)$ and the material's electric susceptibility $\chi(\qb,\omega)$ comprise the whole description of optical media in the framework of Maxwell's equations.

In the realm of quantum theory, the formation of AC current-density waves $\Jbac(\rb, t)$ is the result of photoexcited transitions among electron states in momentum space, as illustrated in Fig.\ \ref{Fig1}(d) [examples of states in real space are shown by dashed curves in Fig.\ \ref{Fig1}(b)]. A photon propagating inside a medium that contains Bloch electron states can be regarded as a coupled (photon+electrons) state described by a Hamiltonian $\hat{H}_{\rm p}+\pb^2/2\me-eV(\rb)+\Hi$ (see supplementary information), where terms from left to right stand for the energy of the free EM field, the electrons in a model periodic ionic potential $V(\rb)$, and the light-matter interaction. The electron states are Bloch waves defined as eigenfunctions of $\pb^2/2\me-eV(r)$, namely $\psi_\jk=\ee^{\ii\kb\cdot\rb} u_\jk(\rb)/\sqrt{\nu}$, where $\nu$ is a normalization volume in 3D or area in 2D, $\kb$ is the Bloch wave vector, and $u_\jk (\rb)$ is a normalized function with the same periodicity as the atomic lattice. Under the perturbation induced by $\Hi$ with a time dependence imposed by the light frequency $\omega$, quantum transitions among electron states [color-filled circles in Fig.\ \ref{Fig1}(b) and (d)] give rise to superposition wave functions $\Psi(\rb,t)=c_1 \psi_{j\kb_1}+c_2 \psi_{j'\kb_2} \ee^{-\ii\omega t}$, where the time-dependent phase difference $\ee^{-\ii\omega t}$ is retained (see supplementary information). The electric current density described by $\Psi(\rb,t)$ is easily evaluated as $\jb=-e\R{\Psi^*\pb\Psi}/\me$, which gives rise to AC and DC components (see supplementary information)

\begin{align}
\jbac  &  \approx  \frac{-e}{\nu\me} \R{c_1^* c_2 \ee^{\ii(\qb\cdot\rb-\omega t)}}\hbar (\kb_1+\kb_2),              \label{jac}\\
\jbdc  &  =        \frac{-e}{\nu\me} \left(\abs{c_1}^2\mean{\pb}_{j\kb_1}+\abs{c_2}^2\mean{\pb}_{j'\kb_2}\right),   \label{jdc}
\end{align}

\noindent respectively. The approximation in Eq.\ (\ref{jac}) is made for nearly free electrons, but our discussion is generally valid for more rigorous results shown below. The AC current in Eq.\ (\ref{jac}) is a propagating wave characterized by a wave vector $\qb=\kb_2-\kb_1$, while the current direction is aligned with the sum of momenta of initial and final electron states $\kb_1+\kb_2$ (see supplementary information). The DC current in Eq.\ (\ref{jdc}), which is proportional to the total momentum of the electron, receives contributions $\mean{\pb}_{j\kb_i}=\bra{\psi_{j\kb_i}}\pb\ket{\psi_{j\kb_i}}$ from both Bloch states $i=1,2$.

The polarization current-density waves $\Jbac(\rb,t)$ entering the Maxwell equations must result from the sum of the AC current waves in Eq.\ (\ref{jac}) for all possible transitions in $\kb$-space [Fig.\ \ref{Fig1}(d)]. For the plasmons and transverse EM waves considered in Fig.\ \ref{Fig1}(b) and (c), where the electric field (in-plane component for plasmons) is parallel ($\Eb_\parallel(\rb,t)$) and perpendicular ($\Eb_\perp(\rb,t)$) to $\qb$, respectively, the AC currents in Eq.\ (\ref{jac}) produced by two transitions symmetrically distributed in $\kb$-space [$\jb_{\rm AC,I}$ and $\jb_{\rm AC,II}$ in Fig.\ \ref{Fig1}(c)] are in-phase or out-of-phase, so that the total current- density waves $\Jb_{\rm AC}(\rb,t)$ are longitudinal or transverse, respectively. In addition, the charge-density wave in the plasmons can also be understood as a quantum superposition, as illustrated in Fig.\ \ref{Fig1}(a) by considering two quantum states with $\kb_{1,2}$ both along $\hat{x}$ (dashed curves, Fig.\ \ref{Fig1}{\it A}), where the charge density appears as a propagating wave of charge density $\rho(x,t)=-e\abs{\Psi(x,t)}^2=-2e\R{c_1^*c_2 \ee^{\ii(q x-\omega t)}}/\nu$ [Fig.\ \ref{Fig1}(a), color scale].

The polarization current-density turns out to be proportional to the optical electric field, and we can write $\Jbac(\rb, t)=-\R{\ii\omega\chi(\qb,\omega)\Eb(\omega)\ee^{\ii(\qb\cdot\rb-\omega t)}}$, which defines the electric susceptibility $\chi(\qb,\omega)$ of the optical medium. Based on a more rigorous form of $\jbac$ beyond that of Eq.\ (\ref{jac}), and with a minimal coupling interaction Hamiltonian $\Hi=e(\pb\cdot\Ab+\Ab\cdot\pb)/2\me-e\phi$, this procedure reproduces the well-established result \cite{Wooten2013} (see supplementary information)

\begin{align}
\chi_{\perp,\parallel}(\qb,\omega) &= -\frac{\omega_{\rm p}^2}{\omega^2} - \frac{\omega_{\rm p}^2}{\omega^2} \frac{2\hbar}{M}\times        \nonumber\\
                 &\sum_{jj',\kb} |K_{\perp,\parallel}|^2 \frac{\nfdif}{\nomde}       ,\label{Chi}
\end{align}

\noindent where $\chi_{\perp}$ and $\chi_{\parallel}$ stand for the response to transverse ($\Eb_\perp$) and longitudinal ($\Eb_\parallel$) electric field components, we include the contribution of the paramagnetic current in the first term, $\omega_{\rm p}=\sqrt{N e^2/\nu\varepsilon_0\me}$ is the plasmon frequency, $N$ is the total number of charge carriers, $M=N\me$ is the total mass of the electrons in the system, $K_\perp$ and $K_\parallel$ are components of the transition matrix element $\Kb=\bra{u_\jk}\kb+\qb/2-\ii\nabla \ket{u_\jpkq}/\nu$ perpendicular and parallel to $\qb$, and $f_\jk$ and $\hbar\omega_\jk$ are the population and energy of state $\psi_\jk$. In the local limit ($q\rightarrow0$), the contributions of intra- and interband transitions reduce to the Drude model and Lorentz model \cite{Adler1962}, respectively.

\subsection{Theory of material momentum in optical media}

The standard description of the electric susceptibility provided by Eq.\ (\ref{Chi}), derived in a quantum approach and meant to capture the AC response of the medium illustrated by Eq.\ (\ref{jac}), is commonly used in combination with Maxwell's equations for a self-consistent treatment of the fields. However, the DC current in Eq.\ (\ref{jdc}), which is proportion to the increase in electron momentum, is not present in Maxwell's equations. In analogy to the derivation of Eq.\ (\ref{Chi}), by summing over all possible transitions, we can find the total momentum gained by the electrons $\ge$ due to polarization in the medium. For a transverse EM wave as shown in Fig.\ \ref{Fig1}(c), the resulting final electron momentum is explicitly given by (see supplementary information)

\begin{align}
\gb_{\rm ele,mat}^\perp= -&\frac{\varepsilon_0\abs{E_\perp(\omega)}^2}{2} \frac{\omega_p^2}{\omega^2} \frac{1}{M}
\sum_{jj',\kb}\abs{K_\perp}^2\; \Delta \gb \nonumber\\
&\times\frac{\nfdif}{(\nomde)^2}  ,\label{gT}
\end{align}

\noindent and for the longitudinal electric field component [Fig.\ \ref{Fig1}(b)], the electron momentum reduces to

\begin{align}
\gb_{\rm ele,mat}^\parallel=&-\frac{e^2|E_\parallel(\omega)|^2}{2\hbar^2 q^2\nu}
\sum_{jj',\kb}  |\langle \psi_\jk|\ee^{-\ii \qb\cdot\rb}|\psi_\jpkq\rangle|^2 \Delta \gb   \nonumber \\
&\times \frac{\nfdif}{(\nomde)^2}        ,\label{gL}
\end{align}

\noindent where $\Delta\gb=\mean{\pb}_\jpkq-\mean{\pb}_\jk$ is evaluated for Bloch states. Likewise, the total momentum $\gbm$ transferred to the electrons and the lattice during the emergence of polarization is also expressed by Eqs.\ (\ref{gT}) and (\ref{gL}) by simply redefining $\Delta\gb=\hbar\qb$.

To derive $\gbe$ in Eqs.\ (\ref{gT}) and (\ref{gL}), we have to sum the gauge-independent kinetic momentum of all electrons, $\bra{\Psi} \pb+e\Ab(\rb,t) \ket{\Psi}$, evaluated in the final superposition state $\Psi$ (see previous section). Without loss of generality, we can chose the Coulomb gauge, in which the term $e\Ab(\rb,t)$ has no contribution, so that the change in electron momentum associated with each electron transition in Eqs.\ (\ref{gT}) and (\ref{gL}) is $\Delta\gb=\mean{\pb}_\jpkq-\mean{\pb}_\jk$. In contrast, during each electron transition, the momentum transferred from the perturbative field to the whole material is $\Delta\gb=\hbar\qb$, which leads to the expression for $\gm$ in Eqs.\ (\ref{gT}) and (\ref{gL}).

In general, during the emergence of polarization, the material momentum is imparted to both electrons and lattice, $\gm=\ge+ \gl$. For free electron systems without a lattice potential, we have $\mean{\pb}_\jk=\hbar\kb$ and thus $\Delta \gb=\hbar\qb$, so according to Eqs.\ (\ref{gT}) and (\ref{gL}), we can conclude $\gm=\ge$, which is an intuitive result because here the material momentum is entirely supported by the electrons ($g_{\rm lat}=0$). Insulating media constitute another extreme scenario in which both the DC current and $\ge$ are absent (see below), so the material momentum only manifests in lattice motion (i.e., $\gm=\gl$).

\subsection{Revisiting the AM controversy}

The material momentum $\gm$ derived above in fact reveals the hidden momentum focused in the context of the AM controversy. With the result obtained for $g^\perp_{\rm mat}$, we can explicitly calculate the ratio of momenta distributed in the EM field and matter for a transverse EM plane wave as shown in Fig.\ \ref{Fig1}(c). Considering the field momentum of a transverse EM plane wave as obtained from the Maxwell equations, $\gf=\varepsilon_0|E_\perp(\omega)|^2 q/2 \omega$ \cite{Jackson2007}, we readily find an equation that is in agreement with previous classical resolutions to the AM controversy \cite{Barnett2010,Mansuripur2010,Kemp2011}

\begin{align}
\frac{g^\perp_{\rm mat}}{\gf}=\chi_\perp(\omega)+\frac{\omega}{2} \frac{\dd\chi_\perp(\omega)}{\dd\omega}.   \label{ratio}
\end{align}

\noindent According to Minkowski's expression, the momentum of an individual photon in an EM plane wave is $\gM=n_{\rm p}\hbar q_0$, where $n_{\rm p}(\omega)=\sqrt{1+\chi_\perp(\omega)}$ is the phase index, and $\chi_\perp(\omega)$ is determined by Eq.\ (\ref{Chi}). For a dispersive material, Abraham's form of the photon momentum is $\gA=\hbar q_0/n_{\rm g}$, where $n_{\rm g}(\omega)=\dd(\omega n_{\rm p})/\dd\omega$ is the group index. Taking these relations into consideration, the ratio in Eq.\ (\ref{ratio}) reduces to $g_{\rm ele}^\perp/\gf=(\gM-\gA)/\gA$, which is consistent with previous solutions to the AM controversy that the Abraham form of the photon momentum only accounts for the EM field, while Minkowski's expression includes the momentum imparted to the material, in agreement with the conclusion extracted in previous works.

Our theory explains clearly the process of momentum transfer in a lossless material. Since our study focuses on lossless media, the phenomenological damping rate $\gamma$ that usually appears in the denominator of Eq.\ (\ref{Chi}) as $\omega_\jpkq-\omega_\jk-\omega+i\gamma$ (see {\it SI Appendix}, section S3) and similarly in Eqs.\ (\ref{gT}) and (\ref{gL}), is ignored ($\gamma\rightarrow 0$). However, we must retain a nonzero (although infinitesimal) value of $\gamma$, needed to prevent a divergent heating of the electron system, as explained in previous studies on quantum Floquet systems \cite{Prosen1998,Prosen1999}. Regarding the temporal response of the optical medium, $1/\gamma$ characterizes a timescale in which momentum is transferred to the material. In the dissipationless limit $\gamma\rightarrow 0$, the momentum transfer process thus takes an infinitely long time ($\sim1/\gamma$) before the system reaches a steady state, and the total momentum transferred to the material is convergent and determined by Eqs.\ (\ref{gT}) and (\ref{ratio}). The momentum transfer process here discussed is further corroborated through simulations in a subsequent section.

We also highlight the importance of the frame of reference for this discussion, as also revealed in previous studies \cite{Wang2015}: if the frame is chosen as we adopt in this study, in which the material (electron and lattice in thermal equilibrium) is initially motionless before the EM field is applied then after the steady distribution of polarization in the media is established, the total momentum should be consistent with the Minkowski formalism, and the momentum gained by the material is just the difference between Abraham and Minkowski formalisms, as revealed by $\ge/\gf=(\gM-\gA)/\gA$. However, if one assumes a steady EM wave propagation in an insulating medium with the frame of reference fixed on the lattice, the material momentum $\gm$ is absent in this scenario, since $\gl=0$ and $\ge=0$ (absence of $\Jdc$ in an insulator, see below), so that the total momentum should be $\gt = \gf$, as obtained from the Abraham formalism. In a general situation of a conducting medium (not totally free electrons) with the frame of reference fixed on the lattice, the lattice momentum is still absent (i.e., $\gl=0$), while neither the Minkowski nor the Abraham momentum accounts for the total momentum. We note that we neglect relativistic corrections for the transformation between the two frames due to the small lattice velocities under consideration.

\subsection{Electron momentum in 2D electron systems}
In free electron systems, the material momentum is equal to the electron momentum $\gm=\ge$. Here, we focus on the role of 2D plasmons in a 2DEG [Fig.\ \ref{Fig1}(b)] as an example to illustrate $g_{\rm elel}^\parallel$ (hereafter, scripts $\perp$ and $\parallel$ are omitted for simplicity without ambiguity) in a free electron system. The results are shown in Fig.\ \ref{Fig2}. Directly applying Eqs.\ (\ref{Chi}) and (\ref{gL}) to the 2DEG, we can immediately find the effective refractive index $n_{\rm eff}$ of the plasmon mode [Fig.\ \ref{Fig2}(a)] and the momentum of the constituent electrons, plotted in Fig.\ \ref{Fig2}(b), where it is normalized to $\gf$. The latter is calculated for 2DEG by integrating $E_y H_z/2c^2$ along $\hat{z}$ (see Appendix). Interestingly, using the knowledge acquired on the AM controversy in the previous section, we can also try to estimate the material momentum as $\gm/\gf=(\gM-\gA)/\gA$. For plasmons in the 2DEG (and in general for similar waveguiding modes), both the Minkowski and Abraham momenta, $\gM=n_{\rm eff}\hbar\omega$ and $\gA=\hbar\omega/n_{\rm g}$, respectively, should be defined in terms of $n_{\rm eff}$ and the group index $n_{\rm g}=\dd(\omega n_{\rm eff})/\dd\omega$.

In Fig.\ \ref{Fig2}(b), we compare rigorous numerical results of $\gm/\gf=\ge/\gf$ (solid curve) with the estimate obtained from the AM formalism (dashed curve). The latter is in good agreement with our quantum theory over a wide frequency range. In fact, we can analytically prove that in the local limit the electric susceptibility in Eq.\ (\ref{Chi}) reduces to the Drude model and the electron momentum in Eq.\ (\ref{gL}) is exactly equivalent to the AM estimate (see {\it SI Appendix}, section S5). For higher frequencies, plasmons in the 2DEG show a large $\nef(\omega)$ and strong confinement, where nonlocal corrections are not negligible (e.g., the plasmon wave vector $q$ is no longer negligible compared with the Fermi wave vector $\kf$), and as a result of the quantum theory here presented deviates from the AM estimate [see Fig.\ \ref{Fig2}(b)].

Our quantum model can also be generalized to investigate the momentum of graphene plasmons \cite{Hwang2007,Chen2012,Fei2012}. The plasmon dispersion curves [Fig.\ \ref{Fig2}(c)] and EM momentum $\gf$ in freestanding graphene can be obtained from the above equations for the 2DEG simply by using the electric susceptibility of graphene, which is obtained in a similar way as Eq.\ (\ref{Chi}). We note that Dirac electrons in graphene are not truly free, as they display conical rather than parabolic dispersion. Using Dirac spinors to represente the electron wave functions \cite{Neto2009}, and taking into account both intra- and interband transitions, we find the electron and material momenta from Eq.\ (\ref{gL}) (see {\it SI Appendix}, section S6). The results are shown in Fig.\ \ref{Fig2}(d) under the same conditions as in Fig.\ \ref{Fig2}(c). In graphene plasmon, the electron momentum $\ge$ is also associated with a DC electric current as already observed in experiment \cite{Karch2010}, and the material momentum $\gm$ can be estimated according to the above discussion of the AM controversy [dashed curve, Fig.\ \ref{Fig2}]. Interestingly, the electron momentum $\ge$ in graphene is much larger then the material momentum $\gm$, which implies that the lattice momentum should be antiparallel with respect to the photonic wave vector according to $\gm=\ge+\gl$, so we conclude that the excitation of graphene plasmons results in a pulling force on the graphene lattice.

\subsection{Electron momentum and band structure}

In a more realistic description, we consider media with multiple electron bands, so the electron momentum $g_{\rm ele}$ expressed in Eqs.\ (\ref{gT}) and (\ref{gL}) includes detailed information of the electron group velocity of electrons, as shown in the definition of $\Delta g$. To understand these effects, we study a model consisting of an infinite one-dimensional atomic chain under illumination by an EM plane wave, as illustrated in Fig.\ \ref{Fig3}(a). The wave vector component of the plane wave parallel to the chain is the photonic wave vector of interest, which is also denoted as $\qb$ [see dashed black arrow in Fig.\ \ref{Fig3}(a)]. The optical response of the chain to the electric field components parallel ($\Eb_\parallel$) and perpendicular ($\Eb_\perp$) to $\qb$, which are uncoupled for a pertubative field, can be regarded as simplified one-dimensional analogs of the longitudinal and transverse wave propagation illustrated in Fig.\ \ref{Fig1}(b) and (c), respectively. Since the change in lattice velocity due to the momentum $\gl$ transferred to the lattice is much smaller than the electron drift velocity, $\ge$ can be considered to be invariant in the two aforementioned frames of reference, so we can adopt the one fixed on the lattice for the following discussion.

We consider a tight-binding nearest-neighbors model of the atomic chain, with each atom containing an s-type orbital [blue circle, Fig.\ \ref{Fig3}(a)] and two degenerate p-type orbitals [red ellipse, Fig.\ \ref{Fig3}(a)]. The latter are oriented either parallel (p$_\parallel$) or perpendicular (p$_\perp$) to the chain. We only consider electron orbital couplings between adjacent s-orbitals (s-s coupling) and adjacent p-orbitals aligned in the same direction (i.e., p$_\parallel$-p$_\parallel$ and p$_\perp$-p$_\perp$ couplings; s-p$_\parallel$ coupling is thus ignored). Diagonalization of the tight-binding Hamiltonian results in Bloch states $\psi_{jk}$ ($j={\rm s,p_\parallel,p_\perp}$) expanding to an energy band $\hbar\omega_{jk}=\hbar\omega_j^0-2t_j\cos(ka)$, where $k$ is the Bloch wave vector, $\hbar\omega_j^0$ are the uncoupled electron energy levels, and $t_j$ are the hopping energing for corresponding each type of coupling $j$ (i.e., s-s, etc.). The band structure of the atomic chain is shown in Fig.\ \ref{Fig3}(b), where we assume $\hbar\omega_{\rm s}^0=E_0$, $\hbar\omega_{\rm p}^0=5E_0$, $t_{\rm s}=0.5E_0$, $t_{\rm p\parallel}=-0.8E_0$ and $t_{\rm p\perp}=0.4E_0$, all normalized to a characteristic energy $E_0$. The sign of $t_j$ and thus also the curvature of the band dispersion [see Fig.\ \ref{Fig3}(b)] are determined by the parity of the involved atomic orbitals. In the simulations and calculations that follow (see Appendix), we assume a chain of $L=101$ atoms with periodic boundary conditions and a lattice constant $a$. The projection of the photonic wavelength along the chain gives $2\pi/q=\ell a$. Because $2\pi/q \gg a$, we can take $L=\ell$; this choice does not affect the results and conclusions extracted below. In fact, with $L=\ell$, the topology of the system is equivalent to an atomic ring under illumination by circularly polarized light.

For the excitation with the electric field component perpendicular to the chain $\Eb_\perp$, only interband transitions between the s-band and p$_\perp$-band are allowed due to the symmetry of the system. For a photon energy smaller than the band gap energy $\Delta E=1.4E_0$ [see black arrow and light grey area in Fig.\ \ref{Fig3}(b)], one might naively conclude that the photon cannot induce interband transitions, as those quoted to explain the photocarrier generations in photogalvanic effect in a semiconductor. However, we argue that this below-band gap light frequency can also cause weakly absorptive interband electron transitions, which eventually populate the upper electron bands in the dissipationless limit. To demonstrate this, we perform a time-dependent simulation with an insulating chain modeled in Fig.\ \ref{Fig3}(a) and (b) ($\Ef=2E_0$), excited by the electric field component $\Eb_\perp$ with a photon energy $\hbar\omega=E_0$ smaller than the band gap energy. In the simulation, a vanishingly weak damping mechanism (see Appendix), which is necessary as explained above, is captured by a phenomenological small damping rate $\gamma$. After a long enough simulation time ($\gg 1/\gamma$), we observe a nonzero steady population in the upper conduction band, as shown in Fig.\ \ref{Fig3}(c), which is independent on the value of $\gamma$ we chose. These interband transitions seem to be at odds with the assumption $\hbar\omega<\Delta E$. However, one must keep in mind that the conduction band in an insulator also contributes to the optical response when $\hbar\omega<\Delta E$, as can be seen in Eqs.\ (\ref{Chi}), (\ref{gT}) and (\ref{gL}). Without inclusion of these interband transitions, the medium are not polarizable, implying the unphysical result $\chi(\omega)=0$ and $n(\omega)=1$ inside any transparent material.

For an ideal insulator with Fermi level $\Ef$ located in the band gap as in Fig.\ \ref{Fig3}(c), the steady populations established under light excitation in both the conduction and valence bands are symmetric in momentum space, leading to a zero DC current $\Jdc$, which is consistent with the insulating character of the medium. In fact, the vanishing of $\ge$ for filled bands can be rigorously proved from Eqs.\ (\ref{gT}) and (\ref{gL}), considering the relation $\mean{\pb}_\jk=\me\nabla_\kb \omega_\jk$ (see {\it SI Appendix}, section S2) and the periodicity of $\omega_\jk$. However, for a medium with partially filled bands, a DC current can arise due to weakly absorptive interband transitions, as confirmed in Fig.\ \ref{Fig3}(d). Figure\ \ref{Fig3}(d) shows the time evolution of the instantaneous space-averaged electric current $\mean{J}$ in an atomic chain, where the Fermi energy $\Ef=1.6E_0$ leaves the s-band partially filled [see Fig.\ \ref{Fig3}(b)] and illumination with a field component $\Eb_\perp$ is assumed. In all of the simulations with different damping rates $\gamma$ plotted in Fig.\ \ref{Fig3}(d), the system is eventually relaxed to the same steady state, as shown from the same final steady DC currents $J_{\rm DC}$. This independence of $J_{\rm DC}$ on $\gamma$ demonstrates that the electron momentum $g_{\rm ele}$ (and similarly other material-associated momenta including $g_{\rm mat}$ and $g_{\rm lat}$) as determined by Eqs.\ (\ref{gT}) and (\ref{gL}) for $\hbar\omega<\Delta E$ are intrinsic. The electron momentum $g_{\rm ele}$ is transferred from the perturbative EM field to the medium over a time scale $1/\gamma$, and a reduction in $\gamma$ simply leads to longer relaxation time needed by the simulations to reach the same final steady state. The intrinsic nature of $g_{\rm ele}$ for a photon energy $\hbar\omega< \Delta E$ is also implied by the convergence of Eqs. (\ref{gT}) and (\ref{gL}) for $\hbar\omega<\Delta E$. In contrast, Eqs.\ (\ref{gT}) and \ref{gL} are divergent for a photon energy $\hbar\omega>\Delta E$, similar to what happens in the photogalvanic effect. The latter is thus highly absorptive, so a finite $\gamma$ in Eqs. (\ref{gT}) and (\ref{gL}) is necessary to avoid singularities, resulting in a photocurrent that depends on the actual value of $\gamma$.

More strikingly, we also observe that the DC current $\Jdc$ can be directed along the wave vector $\qb$ for certain ranges of the Fermi level $\Ef$, indicating an optical pulling effect acting on the electrons. We simulate the steady DC current $\Jdc$ in the atomic chain for different values of $\Ef$ under illumination with $\Eb_\perp$ [dots in Fig.\ \ref{Fig3}(e)]. In fact, $\Jdc$ can also be directly calculated from Eq.\ (\ref{gT}) [curves, Fig.\ \ref{Fig3}(e)], resulting in remarkably good agreement with the simulation results. In Fig.\ \ref{Fig3}(e), the direction of $\Jdc$ is oriented along $\qb$ when $\Ef$ is placed within the s-band (see red-filled area). In this scenario, the electron momentum $\ge$ is antiparallel with $\qb$, implying that the electrons in the chain are pulled backward by the incident light wave.

For a field polarization $\Eb_\parallel$ parallel to the chain, in addition to intraband transitions between the s- and p$_{\parallel}$-bands (here, we also have $\hbar\omega=E_0<\Delta E$), the symmetry of the system further enbables intraband transitions, which in fact dominate the material response and eventually result in a one-dimensional plasmon wave (see Appendix). Similar to Fig.\ \ref{Fig3}(e), we plot in Fig.\ \ref{Fig3}(f) the current $J_{\rm DC}$ in the atomic chain with different values of $\Ef$ for excitation by $\Eb_\parallel$. Our simulations are again in good agreement with Eq.\ (\ref{gL}). For certain ranges of $\Ef$ within both s- and p$_{\parallel}$-bands, we again observe $J_{\rm DC}$ along the wave vector $\qb$ (see red-filled areas), demonstrating the existence of an electron pulling effect associated with plasmon waves. In fact, a similar phenomenon has been observed in a recent experiment on a thin metal film \cite{Strait2019}, which is still regarded as a perplexing result. A theoretical explanation for this experimental observation is yet absent, since it is widely accepted that plasmons in a conductor can be adequately described through a free electron gas model, in which the momentum transferred to the electrons should always push them forward, as shown in Fig.\ \ref{Fig2}. However, the theory and results here presented reveal that a proper description of $\ge$ requires richer information on the electron band structure. For instance, in the local limit $q\rightarrow 0$, Eq.\ (\ref{gL}) for the one-dimensional system includes $\Delta g \propto \nabla^2_\kb \omega_{j\kb}$, while the AC optical response in Eq.\ (\ref{Chi}) only involves $\omega_{j\kb+\qb}- \omega_{j\kb} \propto\nabla_\kb \omega_\jk$. Nonetheless, the overall momentum conservation among EM field, electrons and lattice should be satisfied, with the material momentum $\ge$ determined according to our previous discussion on the AM controversy.

\begin{figure}[h!]
\centering
\includegraphics[width= 0.8\columnwidth]{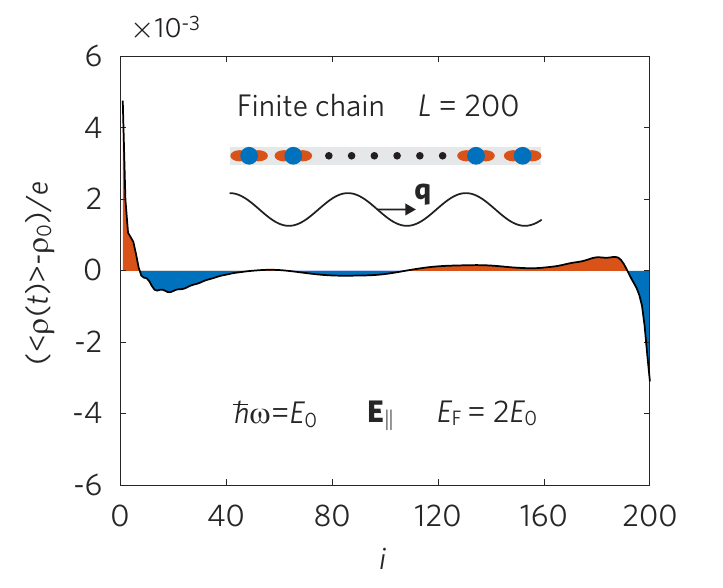}
\caption{ Charge distribution in an insulating 1D atomic chain under light excitation. The atomic chain contains $200$ atoms, and all other parameters are same as in Fig.\ \ref{Fig3}. The lower s-band is filled and the Fermi level is $\Ef=2 E_0$, so the unperturbed electron charge density is $\rho_0=-e$. The projection of the photonic wavelength along the chain is $2\pi/q=80a$. A longitudinal field $\Eb_\parallel$ is considered for illustration, and the photon energy is $\hbar\omega=E_0$, which is below the band gap energy. In the insulating chain, a permanent electric dipole is observed at the two edges, thus explaining the emergence of an optical pressure on material boundaries.
} \label{Fig4}
\end{figure}

\subsection{Optical pressure and boundary charges}

Previous classical results conclude two possible mechanisms for the emergence of optical pressure in material media. In the first mechanism, light absorption directly transfers momentum to the material. In our theory, for the lossless media under consideration, this momentum transfer process takes place over an infinitely long time scale $\sim1/\gamma$ in the dissipationless limit ($\gamma\rightarrow 0$), so the resulting optical pressure is infinitesimal, provided that the amount of momentum imparted to the material is finite, as determined by Eqs.\ (\ref{gL}) and (\ref{gT}).

A second mechanism refers to optical pressure on material boundaries, such as the optical force predicted by Maxwell's equations for optical transmission through a lossless finite dielectric slab. Our study so far is focused on infinite media, so in order to investigate this mechanism specific of finite media, we consider a tight-binding model of an insulating atomic chain, in which the chain is truncated to contain a finite number of atoms $L=200$ (see inset in Fig.\ \ref{Fig4}). We perform time-dependent simulations with the model atomic chain excited by a $\Eb_\parallel$ light wave (note that intraband transitions are absent), assuming a photonic wave vector $q=2\pi/80a$ along the chain. After the system reaches a steady state, we continue the simulation for a sufficiently long time to find the time-averaged charge distribution shown in Fig.\ \ref{Fig4}, where time-harmonic charge oscillations are averaged out. Interestingly, despite the insulating character of the atomic chain, we observe in Fig.\ \ref{Fig4} a depletion and accumulation of electron density at the front ($i=1$) and rear ($i=200$) boundaries, respectively. Permanent electric dipole moments parallel to the wave vector are thus established near the two boundaries. The emergence of these boundary charges is due to translational symmetry breaking. The boundary electric charges and dipole moments revealed in Fig.\ \ref{Fig4} provide a microscopic explanation for the emergence of optical pressure on material boundaries. As shown in Fig.\ \ref{Fig4}, the electron distribution near the boundaries is directly distorted in response to the propagating EM wave. Due to the net accumulation of charge resulting from this electron density distortion, the ionic lattice, which should be relatively inert to the oscillating EM field, eventually receives an optical pressure from the disturbed electrons via Coulomb interaction.

\section{Conclusion}

In summary, with this work we clearly disentangle the different types of momenta associated with photon propagation in infinite extended and lossless media, where we decompose the total momentum $\gt$ into components associated with the EM field $\gf$, the electrons $\ge$ and the lattice $\ge$. In addition, $\ge$ and $\gl$ compose the momentum accommodated in the material $\gm=\ge+\gl$. We then develop a theory that leads to explicit results for these types of momenta. In particular, the result for $\gm$ obtained here from first principles is consistent with previous solutions to the AM controversy based on classical theory. Beyond the classical AM controversy, our theory can adequately address the electron momentum $\ge$ associated with light propagation in a medium, which manifests in an DC current. We find that $\ge$ depends on the electron band dispersion of the material, and our calculations reveal that under certain circumstances the electrons are pulled backward by the propagating EM wave. Our study also offers a new microscopic interpretation for the optical force acting on finite systems, predicting that spatially nonuniform charges are accumulated at the boundaries of the materials and subsequently exert a force on the atomic lattice via Coulomb interaction.

With new developments in various subdisciplines of physics, the concept of optical media has extended to a much broader range than what previous studies on the AM controversy are capable of addressing. It is important to understand the momentum associated with photon-coupled states in relevant types of materials supporting unconventional excitations, such as material involving quantum nonlocal effects, as well as microscopic materials without a well-defined refractive index, or materials with novel topological properties. Since our theory is based on first principle, it can be employed for that purpose, as inferred from the discussions presented above for strongly confined 2D plasmons and microscopic atomic chains.

The DC current resulting from the electron momentum $\ge$ here investigated is intrinsic and physically observable, so it could be measured through the resulting static magnetic field or electric voltage that it generates. The atomic chain in Fig.\ \ref{Fig3} with a periodic boundary condition in our study is equivalent to an atomic ring illuminated by circularly polarized light, so we conclude the existence of DC current loops, when circularly polarized light carrying angular momentum are propagating in a medium or illuminating a small particle. Such DC current loops can give rise to a measurable static magnetic field, which is in fact the inverse Faraday effect. More interestingly, we predict that the electrons in these current loops could acquire angular momentum that is antiparallel the the photon angular momentum under certain conditions, a phenomenon that can be regarded as a negative inverse Faraday effect \cite{Van1965}.

We also note that the intrinsic DC current and the optical force acting on the boundaries of a medium are possibly related to the topological physics of dynamically pumped systems. The light propagation in a crystal medium here investigated in the adiabatic limit $\omega\rightarrow 0$ is in fact an instance of the type of systems studied in the context of the Thouless quantum pump \cite{Thouless1983} in topological physics. The latter predicts a quantized charge transport in a crystal driven by a moving wave-shaped potential in the adiabatic limit, giving rise to a topological quantum Hall effect in some synthetic dimension. The nonadiabatic generalization of the Thouless quantum pump, dealing with a crystal pumped by a fast moving wave-shaped potential similar to that considered in Fig.\ \ref{Fig3}, is still a frontier in the field of quantum pumped systems \cite{Wang2013,Privitera2018}, where the nonadiabatic Chern number should be defined by the matrix in Eq.\ (\ref{Chi}). In our study, the boundary charges and electric dipoles shown in Fig.\ \ref{Fig4} appear to have a resemblance with topological edge modes. In fact, accumulation of charges and electric dipole moments on the boundary of a finite system, with the internal bulk region remaining neutral, is indeed a consequence of topology according to the modern theory of polarization \cite{Resta1994}. Consequently, the present results are potentially instructive for future studies on the momentum and topology in nonadiabatical quantum pumped systems.

\appendix

\section{Derivation of Eq.\ (\ref{ratio})}

For transverse EM waves, the derivative of the electric susceptibility for monochromatic light of frequency $\omega$ in Eq.\ (\ref{Chi}) yields

\begin{align}
\frac{\omega}{2}\frac{\dd\chi_\perp}{\dd\omega} = \frac{\omega_{\rm p}^2}{\omega^2} + &\frac{\omega_{\rm p}^2}{\omega^2} \frac{2\hbar}{M}\times        \nonumber\\
                 &\sum_{jj',\kb} \abs{K_\perp}^2 \frac{\nfdif}{\nomde}       \nonumber\\
                 -\frac{\omega_{\rm p}^2}{\omega} \frac{\hbar}{M}&\sum_{jj',\kb} \abs{K_\perp}^2 \frac{\nfdif}{(\nomde)^2},       \nonumber
\end{align}

\noindent where the first two terms on the right-hand side are simply $-\chi_\perp(\qb,\omega)$. Using Eq.\ (\ref{gT}) and taking into account $\gf=\varepsilon_0 E_y^2/2c$, the remaining third term in the above expression is equal to $g_{\rm ele}^\perp/\gf$, therefore proving Eq.\ (\ref{ratio}).

\section{Dispersion and field momentum of 2D plasmons}
The plasmon modes considered in the 2D electron systems are transverse magnetic (TM) waves. For the 2D plasmons on a free-standing single-layer of 2DEG (Fig.\ \ref{Fig2}), all field components decay exponentially along the out-of-plane direction $\hat{z}$ away from the film. The amplitudes of waves are determined from the boundary conditions imposed from Maxwell's equations, which further leads to the well-known dispersion relation of 2D plasmons $q_z=-2/\chi_\parallel(q,\omega)$, where $q_z=\sqrt{q^2-q_0^2}$, and $q_0$ is the vacuum wave vector. In our calculations, the electric susceptibility $\chi_\parallel(q,\omega)$ in Eq.\ (\ref{Chi}) is used for 2DEGs [Fig.\ \ref{Fig3}(a) and (b)]. Finally, we define the effective refractive index as $\nef(\omega)=qc/\omega$ [Fig.\ \ref{Fig3}(a)].

In the free standing 2D systems considered in Fig.\ \ref{Fig3}, the field is entirely distributed in the vacuum, where the momentum density is unambiguously given by $E_y(z)H_z(z)/2c^2$. The field distribution is simultaneously obtained when solving the dispersion as indicated above. For single-layer systems, the field decays exponentially in the vacuum on either side, so direct integrating along $\hat{z}$ leads to $\gf=\omega\varepsilon_0 q|E_\parallel|^2/2q_z^3 c^2$, where $E_\parallel$ is the field amplitude on the plane of the 2D system.

\section{Modeling of atomic chain}
In Figs.\ \ref{Fig3} and \ref{Fig4}, the atomic chain is described by a tight-binding model. We assume each atom to contain a s-type orbital and two p-type orbitals [p$_\parallel$ and p$_\perp$, see Fig.\ \ref{Fig3}(a)] with nearest-neighbor electron hopping between adjacent orbitals of the same type (i.e., s-s, p$_\parallel$-p$_\parallel$ and p$_\perp$-p$_\perp$). For simplicity, we neglect the coupling between s- and p$_\parallel$-type orbitals, which is in priciple allowed by orbital symmetry. This approximation does not affect the conclusions of this study. The tight-binding electron Hamiltonian of the chain in the atomic coordinate basis is

\begin{align}
\bra{ji}\hat{H}_{\rm TB}\ket{ji'}= \delta_{i,i'}\hbar\omega_j^0+ \delta_{i,i'\pm1}t_j ,\nonumber
\end{align}

\noindent where the indices run over atomic orbital types $j$ ($j={\rm s,p_\parallel,p_\perp}$) and atomic sites $i$, $\hbar\omega_j^0$ is the the uncoupled $j^{\rm th}$ atomic orbital energy, and $t_j$ is the hopping energy between nearest-neighbor orbitals of $j$ type. The infinite atomic chain in Fig.\ \ref{Fig3} is modelled through a finite number of atoms $N$ by imposog periodic boundary conditions (i.e., $\ket{i+L}=\ket{i}$).

The electron eigenfunction $\ket{jm}$ ($j={\rm s,p_\parallel,p_\perp}$) in the chain is a Bloch wave formed by a linear combination of the $j^{\rm th}$ orbital wave function on each atomic site with a relative phase factor: $\ket{jm}=\sum_i\ee^{\ii k_m x_i}\ket{ji}/\sqrt{L}$, where $k_m=2\pi m/La$ is the electron wave vector, $m$ is an integer in the $- L/2 < m < L/2$ range (assuming $L$ is odd), $x_i=i a$ ($i=1,2...L$) is the coordinate of the $i^{\rm th}$ atom, and $a$ is the lattice constant. This basis set diagonalizes the tight-binding Hamiltonian $\hat{H}_{\rm TB}$ according to $\bra{jm}\hat{H}_{\rm TB}\ket{jm'}=\delta_{m,m'}\hbar\omega_{jm}$, where $\hbar\omega_{jm}=\hbar\omega_j^0-2t_j \cos(2\pi m/L)$ are the eigenenergies that produce the band structure shown in Fig.\ \ref{Fig3}(b).

The atomic chain is illuminated by an EM wave with the electric field acting on the chain written as $\Eb(\rb,t)=\Eb \cos(qx -\omega t)$. We assume the projection of the light wavelength along the chain to be $La$, so $q=2\pi/La$. In the Coulomb Gauge, the transverse and longitudinal electric field components are only related to the vector and scalar potentials, $\Eb_\perp(\rb,t)=-\partial_t \Ab(\rb,t)$ and $\Eb_\parallel(\rb,t)=-\partial_x \phi(\rb,t)$. This allows us to write the potentials $\phi(\rb,t)=eE_\parallel\sin(qx -\omega t)/q$ and $\Ab(\rb,t)_\perp=-eE_\perp\sin(qx -\omega t)/\omega \hat{\bf{e}}_\perp$, where $\hat{\bf{e}}_\perp$ is the unit vector perpendicular to the chain. With these explicit forms of the potentials and the Bloch eigenstate basis $\ket{jm}$, following some straightforward algebra, we can find the matrix elements of the light-matter interaction Hamiltonian $\hat{H}_{\rm I}=e\Ab\cdot\pb/\me-e\phi$. The only nonzero matrix elements are

\begin{align}
\bra{jm} \hat{H}_{\rm I}\ket{j'm-1}& \approx \big[\frac{eE_\parallel}{2\ii q} \delta_{jj'}  +\frac{e \Eb_\parallel\cdot\db_{jj'}}{2}  \label{matrixelement}\\
&+\frac{e\Eb_\perp\cdot\db_{jj'} (\omega^0_{jm}-\omega^0_{j'm'})}{2\omega} \big] \ee^{-\ii\omega t}   \nonumber
\end{align}

\noindent and their Hermitian conjugates, where $\db_{jj'}=\bra{j} \rb \ket{j'}$ are the corresponding transition dipole moments. To find Eq.\ (\ref{matrixelement}), we insert the above definition of Bloch states and assume no overlap between orbitals in different atomic sites. The first term in Eq.\ (\ref{matrixelement}) accounts for intraband transitions (i.e., they are diagonal in band index $j$) induced only by the longitudinal field $\Eb_\parallel$; self-consistent interaction can then give rise to chaing plasmons via these terms. The remaining two terms represent interband transitions under excitations by $\Eb_\perp$ and $\Eb_\parallel$, respectively. Considering $|\db_{jj'}| \sim a$, the ratio of intraband to interband transitions is $\sim\lambda/a$, where $\lambda=2\pi/q$, so that intraband processes should dominate the longitudinal response of a medium. In the simulations of Figs.\ \ref{Fig3} and \ref{Fig4}, we assume a weak percolative EM field, $e\Eb_{\perp,\parallel}\cdot\db_{\rm sp}=0.001E_0$, and $eE_\parallel/2\ii q=0.001L E_0$.

Using the total Hamiltonian $\hat{H}=\hat{H}_{\rm TB}+\hat{H}_{\rm I}$, we can then perform time-dependent simulations for the temporal evolution of the density matrix $\rho$ according to the equation of motion

\begin{align}
\dot{\hat{\rho}}=-i[\hat{H},\den]/\hbar-\gamma(\den-\den^0), \nonumber
\end{align}

\noindent where $\gamma$ is a small phenomenological damping rate (see main text). For simplicity, we assume zero temperature, so the initial density matrix in the absence of illumination is $\den^0_{jm,j'm'}=\delta_{jj'}\delta_{mm'}\theta(\hbar\omega_{jm}-\Ef) $, where $\Theta$ is the step function.

The instantaneous space-averaged electric current plotted in Fig.\ \ref{Fig3}(d) is obtaind as

\begin{align}
\mean{J}=-e\sum_{jm} v_{\rm g}^{jm} \den_{jm,jm},
\end{align}

\noindent where $v_{\rm g}^{jm}=2at_j\sin (mqa)/\hbar$ is the group velocity of the electron Bloch state $\ket{jm}$. The off-diagonal elements of the density matrix are related to the oscillating AC current, which has no contribution to the space average. The instantaneous spatial distribution of electric charges is directly obtained from the density matrix in the coordinate basis set, $\rho(t)=-e\sum_{j}^{\rm s,p}\den_{ji,ji}$, and its time average is made after a sufficiently long time-dependent simulation yielding $\mean{\rho(t)}$ in Fig.\ \ref{Fig4}.

\end{document}